\documentclass[12pt]{iopart}
\usepackage{graphicx}
\begin{document}






\title{A proposed atom interferometry determination of $G$ at $10^{-5}$ using a cold atomic fountain}








\author{G. Rosi$^1$}

\address{$^1$INFN Sezione di Firenze - Dipartimento di Fisica e Astronomia, Universit\`a di Firenze, via Sansone 1, I-50019 Sesto Fiorentino (FI), Italy}







\ead{rosi@fi.infn.it}
\vspace{10pt}

\begin{abstract}
In precision metrology the determination of the Newtonian gravity constant $G$ represents a real problem, since its history is plagued by huge unknown discrepancies between a large number 
of independent experiments.
In this paper we propose a novel experimental setup for measuring $G$ with a relative accuracy of $10^{-5}$ using a standard cold atomic 
fountain and matter wave interferometry. We discuss in details the major sources of systematic errors, 
providing also the expected statistical uncertainty. 
Feasibility of determining $G$ at a level of $10^{-6}$ level is also discussed.
\end{abstract}



\section{Introduction}
The Newtonian gravity constant $G$ can be considered the Mt. Everest of precision measurement science \cite{Quinn2014}. After more than two centuries since the
original determination performed by H. Cavendish, experiments based on a torsion balance or torsion pendulum still provide the values with the lowest degree of uncertainty ($\sim20$ ppm).
However discrepancies of several standard deviations between independent measurements are still present \cite{Quinn2013,Faller2010}. One of the reasons that could explain such situation lies in the inherent difficulty
to take fully under control mechanical influences on the so far employed macroscopic probes. Cold atom interferometry has proven to be a powerful and alternative tool for measuring inertial forces \cite{Varenna2014}. The success of this method relies
on the fact that the atomic probe is a microscopic quantum object in free fall that can be precisely controlled and manipulated through frequency-stabilized laser radiation. These features are essential to identify the systematic errors that have proved elusive in previous experiments.
However, no metrologically significant $G$ values have been produced until now.
The explanation for this lies in the fact that in a $G$ determination the gravitational force probed by the atoms is not uniform over the interferometric region. Therefore, a deep characterization of the atomic sample size, trajectory and temperature is required, placing a limit on the final accuracy of the measurement.
To partially solve the problem, a viable option is to employ a set of dense source masses to create stationary points along the vertical acceleration profile. In this case it is possible to identify
a position $\bar{z}$ of the atomic cloud apogee such as

\begin{equation}
\left.\frac{\partial\phi(z)}{\partial z}\right\vert_{\bar{z}}\simeq0
\end{equation}

where $\phi$ is the phase shift of a single interferometer. Thanks to this strategy, a determination of $G$ at the 150 ppm level has been realized \cite{Rosi2014} using an $^{87}$Rb cold atom
gravity gradiometer, which consists of two vertically displaced, simultaneous Raman Mach-Zehnder interferometers. According to the error budget reported in \cite{Prevedelli2014}, the systematic
uncertainty on $G$ due to an error of 0.1 mm on clouds vertical positions is 5 ppm, while the same error on the cloud vertical size ($\sim6$ mm of diameter) produces a much larger shift of 56 ppm. To improve the accuracy by one order of magnitude, the clouds needs to be enclosed in a volume of $\sim1$ mm$^3$ during the ballistic flight. In principle, this can be done using an ultra-cold atomic source \cite{Kovachy2015}. However, bearing in mind 
the typical gradiometer scheme, it is technically challenging to produce a pair of ultra-cold samples and routinely place them with a spatial resolution below 100 $\mu$m. Towards this purpose, schemes based on interferometers trapped in optical lattices
have been proposed and experimentally demonstrated \cite{Zhang2016}. However, they have not yet reached the required maturity for metrological applications.

An alternative way to overcome this type of geometric limitation with a traditional cold-atom gravity-gradiometer can be found following the method suggested in a recent work of A. Roura \cite{Roura2015}. 
Here, it has been demonstrated that changing the module of the Raman $k_{\mathrm{eff}}$ vector by a certain amount $\Delta k_{\mathrm{eff}}$ at the central $\pi$ pulse of a Mach-Zehnder sequence, a phase shift equivalent to the one induced by 
a \textit{fictitious} gravity gradient $\Gamma^{*}_{zz}$ is generated. In particular the following relation holds:

\begin{equation}
\label{eq1}
\Delta k_{\mathrm{eff}}=-(\Gamma^{*}_{zz}T^2/2)k_{\mathrm{eff}}
\end{equation}

where $T$ is the free evolution time between the central $\pi$ pulse and the $\pi/2$ pulses.
Therefore, with a proper choice of $\Delta k_{\mathrm{eff}}$ we can compensate the \textit{real} gravity gradient $\Gamma_{zz}$ probed by the atoms. This matching condition can be easily found by experimentally requiring

\begin{equation}
\label{eq2}
\Phi=k_{\mathrm{eff}}(\Gamma_{zz}-\Gamma_{zz}^{*})(d+\Delta v_zT)T^2=0
\end{equation}

where $\Phi$ is defined as the difference between the upper and lower interferometer phase. Notably, such zero phase shift condition holds regardless of the distance between the atomic samples $d$ and their differential velocity $\Delta v_z$.

This method can also be used to control the systematics in high precision tests of the equivalence principle with atoms in free fall and, more generally, to measure the average gravity gradient experienced by two atomic clouds independently of their relative distance and velocity. Strictly speaking, a geometry-free gravity gradient determination can be realized when the probed acceleration profile $a(z)$ is perfectly linear, or, more generally, if it is possible
to find a condition

\begin{equation}
\label{eq4}
\frac{a(z_2)-a(z_1)}{z_2-z_1}\simeq\Gamma_{zz}(z_2)\simeq\Gamma_{zz}(z_1)
\end{equation}

where $z_2$ and $z_1$ are the coordinates of the clouds apogee.

When applied to a determination of $G$, this method completely redefines the experimental strategy.

In this paper a novel experiment for measuring $G$ with a relative precision of 10 ppm with atom interferometry is proposed and discussed. It will be shown that the target uncertainty can be easily achieved using a cold atomic
fountain of alkali atoms, a technology widely implemented in several metrological institutes worldwide.

The manuscript is organized as follows: in section 2, a principle scheme of the experiment is briefly described. In section 3 requirements in term of statistical uncertainty and sensitivity are reported.
In section 4  the main systematic effects are quantitatively discussed. Finally, in section 5, conclusions and prospects for an experimental determination of $G$ at a level of $10^{-6}$ are presented.

\section{Principle of operation}

\begin{figure}[h]
\centering
\includegraphics[width=0.7\textwidth]{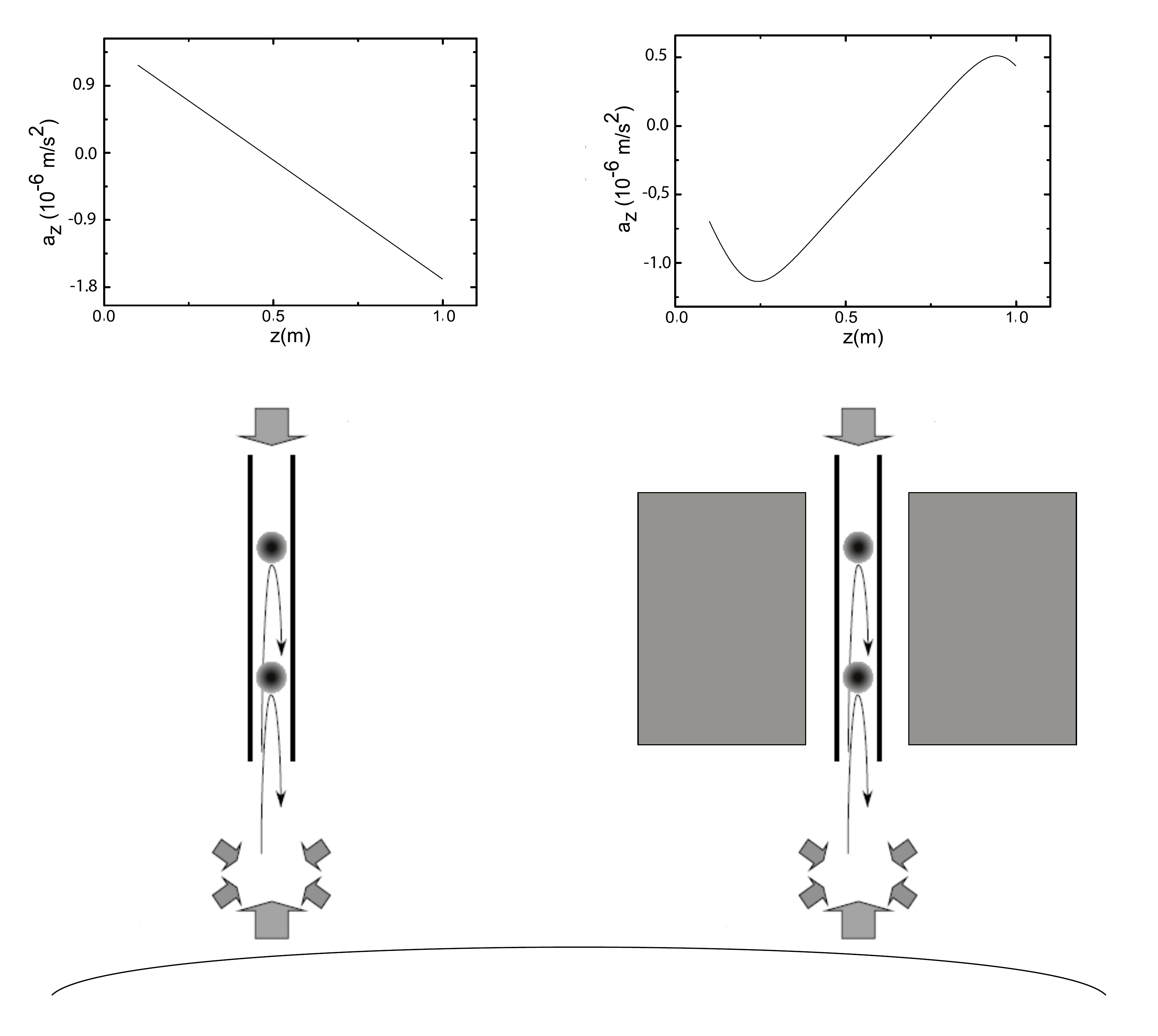}
\caption{\label{fig1}Sketch of the experiment. Two atomic samples are trapped and cooled in a magneto-optical trap (MOT) and sequentially launched towards the interferometric region. A measurement
of the local gravity gradient is performed by Raman interferometry. When local gravity anomalies are far enough (``Far'' configuration, left side), the gravity acceleration profile given by the Earth is almost perfectly linear. The same condition can be also realized by using a proper shaped source mass that surrounds the atomic sensor (``Close'' configuration, right side).
The value of the gravity constant can be retrieved by measuring the corresponding modulation of the gravity gradient.}
\end{figure}

A schematic of the experimental apparatus is reported in figure \ref{fig1}.

Here we consider a vertical gravity gradiometer that consists of a pair of thermal clouds of $^{87}$Rb sequentially launched from a MOT with the standard moving molasses technique and simultaneously interrogated by a sequence of three counter-propagating Raman pulses. A comprehensive description of these well-established methods
can be found in \cite{McGuirk2002,Sorrentino2014} and therefore we will not provide further experimental details about them. The motivation for such a conservative experimental scheme lies in the intention to keep the system as simple as possible. Further efforts to improve the atomic source may be undertaken to push the measurement below the $10^{-5}$ limit.

The value of the gravity gradient experienced by the atoms is alternated between the Earth's gravity gradient (Figure \ref{fig1}, left, ``Far'' configuration) and the combined effect of the Earth and a properly designed source mass (Figure \ref{fig1}, right, ``Close'' configuration). 
From the resulting gravity gradient variation $\Delta\Gamma$, evaluated with the zero phase shift technique, it is possible to retrieve the value of $G$, in a similar way to what was done in \cite{Rosi2014}. Of course, for both the configurations it is crucial to ensure a gravity acceleration profile as linear as possible, in order to actually realize equation \ref{eq4}.

Earth's gravity gradient $\Gamma_{zz}$ is expected to be quite constant as a function of the elevation $h$. According to the free-air correction formula, the second order coefficient is $\simeq h/R_T$ smaller then $\Gamma_{zz}$, where $R_T$ is the Earth's radius \cite{Hackney2003}.
Locally, the acceleration profile can be easily warped by nearby objects and local gravity anomalies. However we can roughly set a requirement on acceleration linearity according to our ability to control the atomic samples vertical coordinate $z$. Let us suppose to have a positional jittering $\delta z\simeq1$ mm and to be in presence of a spherical
anomaly (radius $R$, density contrast $\Delta\rho$) placed below the instrument at a distance $r$. It can be easily found that

\begin{equation}
\frac{\delta\Gamma}{\Gamma_{zz}}=K\Delta\rho\frac{R^3}{r^3}\frac{\delta z}{r}
\end{equation}

with $K\simeq5.4\times10^{-4}$ m$^3$/kg. Imposing $\delta\Gamma/\Gamma_{zz}=10^{-5}$ and taking $\Delta\rho=2\times10^{3}$ kg/m$^3$, we can set some upper limits on the anomaly size. For instance, for $r=1,5,10$ and $50$ m we have $R=0.2,2,4.5$ and $38.6$ m, while $R\simeq r$ at $r=100$ m. We can conclude that the apparatus must be placed sufficiently far from underground structures and
aquifers, while regional scale anomalies can be ignored. A ground-based gravity survey can also help to carefully characterize the area.
It is interesting to point out that the largest mass anomaly in the experiment could be represented by the source mass itself, which must be vertically displaced far enough from the interferometer area in order to  actually realize the ``Far'' configuration.

The source mass should be capable of generating a linear gravity gradient. Moreover, its shape should be as simple as possible, in order to simplify the machining process. A hollow cylinder produces along its vertical axis an acceleration profile with a good degree of linearity (see Figure \ref{fig1}, right), once proper
dimensions and material have been selected. In the following we are going to define such parameters, according to the requirements on statistical and systematic errors.

\section{Statistics}
As mentioned before, the key point of the method relies in determining the zero phase shift condition, which corresponds, according to equation \ref{eq1} and \ref{eq2}, to a frequency jump $\Delta\nu_0=ck_{\mathrm{eff}}\Gamma_{zz}T^2/4\pi$. A naive way to perform such an operation is to measure two gradiometric phases $\Phi(\Delta\nu=0)\equiv\Phi(0)$ and $\Phi(\Delta\nu)\simeq-\Phi(0)$  for each source mass configuration.
It is straightforward to see that

\begin{equation}
\label{eq6}
\Delta\nu_0=\frac{\Phi(0)}{\Phi(0)-\Phi(\Delta\nu)}\Delta\nu
\end{equation}

and

\begin{equation}
\frac{\delta(\Delta\nu_0)}{\Delta\nu_0}=\sqrt{\frac{(\delta\Phi)^2}{\Phi(0)^2}+\frac{2(\delta\Phi)^2}{(\Phi(0)-\Phi(\Delta\nu))^2}+\frac{(\delta\Delta\nu)^2}{\Delta\nu^2}}\simeq\sqrt{\frac{3}{2}}\frac{\delta\Phi}{\Phi(0)}
\end{equation}

The $\delta(\Delta\nu)/\Delta\nu$ term can be neglected since the Raman lasers frequency can be set with a very high degree of precision and stability. To obtain equation \ref{eq6} a linear relation between $\Phi$ and $\Delta\nu$ is assumed. This is strictly true only if $\Delta\nu$ is negligible compared to the detuning $\Delta$ of the Raman lasers from the atomic one photon transition. 
Considering that usually $\Delta\sim1-10$ GHz, this condition can be easily achieved.
Therefore, we can roughly estimate the statical error on $G$ as

\begin{equation}
\left(\frac{\delta G}{G}\right)_{stat}=\frac{\delta(\Delta\nu_0^F-\Delta\nu_0^C)}{\Delta\nu_0^F-\Delta\nu_0^C}=\frac{\sqrt{3}\delta\Phi}{k_{\mathrm{eff}}(\Delta\Gamma)dT^2}
\end{equation}

where the superscripts F and C indicate the Far and Close configuration respectively. In this expression all the relevant experimental parameters are contained: the effective signal $\Delta\Gamma$ is determined by the density of the source mass while its linear dimension is set by $d$ and $T$. The term $\Delta v_zT$ in equation \ref{eq2} can be safely neglected since $d\gg\Delta v_zT$ for our instrument.
Considering tungsten alloy as source mass material ($\Delta\Gamma\simeq6\times10^{-6}$ s$^{-2}$), using standard Raman optics and taking $T=0.243$ s and $d=0.29$ m, we obtain  $\delta\Phi=9$  $\mu$rad for $10$ ppm on $G$. For comparison, in \cite{Rosi2014}, the corresponding $\delta\Phi$ value was $44$ $\mu$rad after 100 hours of integration.
To cover this factor of 5 difference without dramatically increasing the size of the apparatus and/or the integration time, at least two solutions are, in principle, possible. Enhancing $k_{\mathrm{eff}}$ by using multi-photon Bragg pulses is the most evident one. However, at least for thermal clouds,
several factors limiting the contrast for large $T$ are present, in addition to the impossibility of having a convenient internal state mapping \cite{Damico2016}. For Rb, and in general for alkali atoms, one could make use of Raman transitions acting on the $5S\rightarrow6P$ manifold. In this case, with $\lambda=422$ nm, 
a small but metrologically significant gain in sensitivity of 1.8 is ensured. Finally, as it will be shown in the next section, one of the strong point of the presented setup lies in the small systematic contribution due to the cloud atomic size and temperature. As a consequence, atomic flux can be increased
by a factor of ten or even more, for example by relaxing the selectivity in velocity and implementing $m_F=0$ optical pumping. In light of this we can conclude that a statistical error of 10 ppm can be safely achieved in 4-5 days of integration time, once that all the optimizations will be performed. 

\section{Systematics}

In precision measurements, and in particular for a determination of $G$, keeping systematic shifts under control is by far the most demanding task. It is beyond the scope of this work to perform a full \textit{a priori} evaluation of each possible error source. We will focus our attention mostly on the effects produced by the finite dimension of the atomic cloud,
in order to put in evidence the most interesting features of the method. Nevertheless, prior to addressing this topic, it is worth discussing some basic criteria regarding the source mass design. In the following we will consider a monolithic hollow cylinder in order to simplify the discussion. If necessary, an almost equivalent geometry can be realized using smaller cylinders
radially arranged, as done in \cite{Lamporesi2007}.

\subsection{Source Mass}

The vertical acceleration profile $a(z)$ along the axis of a hollow cylinder having a homogeneous density $\rho$, a height $H$, and an inner and outer radius $R_1$ and $R_2$ respectively, can be analytically calculated \cite{Shurr1998}:

\begin{eqnarray}
a(z)&=2\pi\rho G(r_{2-}-r_{2+}+r_{1+}-r_{1-})\nonumber\\
r_{i\pm}&=\sqrt{R_i^2+(\frac{1}{2}H\pm z)^2},\:i=1,2.
\end{eqnarray}

As explained before, the value of $H$ is roughly defined by the gradiometer's vertical extension while $R_1$ and $R_2$ should be chosen in order to produce an acceleration profile with a good degree of linearity. For this purpose it is convenient to expand the previous equation in series:

\begin{eqnarray}
\label{eq3}
\frac{a(z)}{2\pi\rho G}&=H\left(\frac{1}{\sqrt{H^2/4+R_1^2}}-\frac{1}{\sqrt{H^2/4+R_2^2}}\right)z\nonumber\\
&+\frac{H}{2}\left(\frac{R_2^2}{(H^2/4+R_2^2)^{2.5}}-\frac{R_1^2}{(H^2/4+R_1^2)^{2.5}}\right) z^3\nonumber\\
&+ ...
\end{eqnarray}

Here, it is possible to identify several $(R_2-R_1)$ couples that conveniently cancel the cubic term. 

Remarkably, it comes out that having the cubic component small or equal to zero also leads to a reduced sensitivity with respect to the radial coordinate $r$. Indeed, according to \cite{Shurr1998}, it can be
demonstrated that the second derivative of the acceleration with respect to $r$ is half the second derivative of the acceleration with respect to $z$. As a consequence, the advantage of working in the linear region of $a(z)$ rather than in the stationary points is twofold.

Keeping this in mind and taking into account limits due to apparatus size, we can determine suitable test values for the hollow cylinder; in our case, taking into account the interferometer parameters described in the previous section, we choose $R_2=0.556$ m, $R_1=0.120$ m and $H=0.770$ m.
Selecting such large dimensions not only helps in keeping a good $S/N$ ratio but also in relaxing the required accuracy in shape. From equation \ref{eq3} we estimate that an uncertainty of 10 $\mu$m on either height or diameter produces a systematic shift on $G$ of 10 ppm.
A size reduction by a factor 2 or 3 can only worsen the situation. Regarding the material, possible options are lead, tungsten alloy and mercury. The latter is the best in terms of homogeneity, but its large thermal expansion coefficient ($61\times10^{-6}$ K$^{-1}$) requires a temperature stabilization better than 100 mK. Lead is the cheapest,
but also the least dense and the most difficult to machine. Tungsten alloy can be shaped with micrometric precision and its high density ($\rho=18000$ kg/m$^{3}$) is ideal to induce a high gravity gradient. With this material, the total mass of the hollow cylinder amounts to 13 tons. It is worth to point out that such kind of masses
are not new in the field of $G$ metrology \cite{Schlamminger2006}. 

\subsection{Atomic trajectories}

\begin{figure}
\centering
\includegraphics[width=0.7\textwidth]{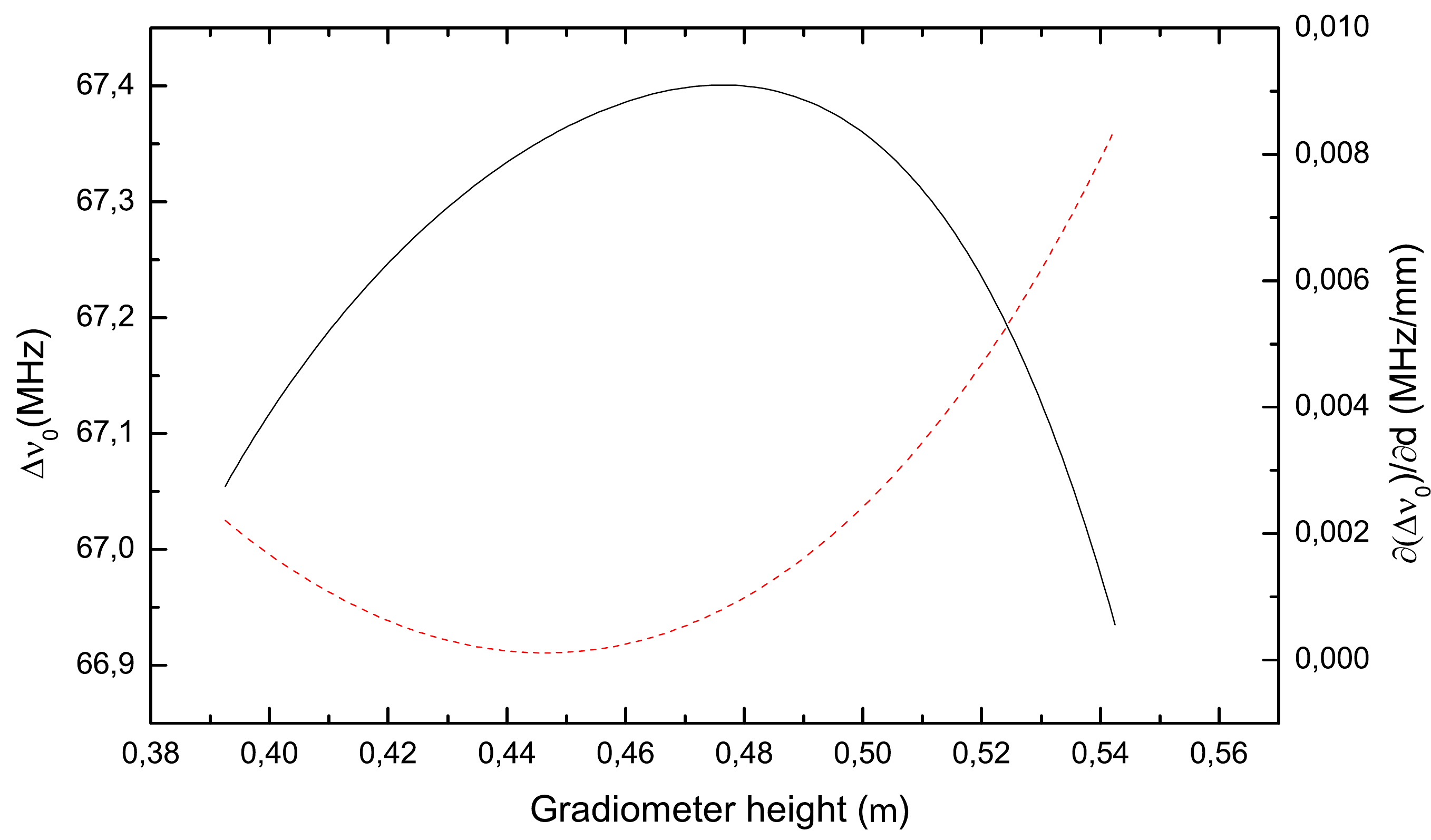}
\caption{\label{deltanu}Black solid line: plot of the simulated zero shift frequency jump $\Delta\nu^{FC}_0$ versus the gradiometer height $z$, defined as the average between the apogee points. Dashed red line: plot of the corresponding $d$ punctual derivative, where $d$ is the vertical distance between interferometers. Source mass lower surface is placed at $z=0$}
\end{figure}

Now that the relevant source mass features have been defined, an evaluation of the expected shifts due to the atomic cloud size can be performed. The first step consists of the optimization of the atomic trajectory along the vertical axis of the system. In Figure \ref{deltanu} the expected value of $\Delta\nu^{FC}_0=\Delta\nu^{F}_0-\Delta\nu^{C}_0$ is plotted versus the gradiometer height $z$, defined as the average vertical coordinates at the apogee.
In addition, the quantity $\partial(\Delta\nu^{FC}_0)/\partial d$ has been evaluated as a function of $z$. A single particle simulation has been employed, using the interferometer parameters defined in section II. We notice that $\Delta\nu^{FC}_0$ cannot be stationary for $z$ and $d$ simultaneously. We choose to fix $z$ such that $\partial(\Delta\nu^{FC}_0)/\partial z=0$ mainly because 
$d$ is easily and accurately measured during the $G$ measurement and therefore less critical.

Since all the single particle parameters have been determined, a Monte Carlo simulation of the experiment has been carried out. An average value for $\Delta\nu^{FC}_0$ has been obtained picking the initial condition according to a given Gaussian position and velocity distribution of the atoms
in the clouds. The parameters entering the simulation are varied and their derivatives calculated to estimate the uncertainty in the measurement of $G$. We considered $10^4$ atoms for each cloud, since increasing such number of a factor 10 does not change significantly the results. Simulation outcomes are summarized in table \ref{tab1}.

\begin{table*}
\caption{\label{tab1} Atomic cloud parameters, corresponding uncertainties and simulated relative shifts on $G$. Uncertainties are quoted as one standard deviation.}
\begin{indented}
\item[]\begin{tabular}{lccc}
\br
&&&Relative uncertainty\\Parameter&Value&Uncertainty&on $G$ (ppm) \\
\mr
Barycentres height (apogee average) &684 mm& 1 mm & $<1$\\
Barycentres vertical distance &290 mm& 0.2 mm & $2.6$\\
Barycentres radial position &0 mm& 2 mm &  $<1$\\
Barycentres radial velocity &0 mm/s& 2 mm/s & $<1$\\
Clouds vertical size (at $\pi/2$) &3 mm& 0.3 mm & $1.7$\\
Clouds horizontal size (at $\pi/2$) &3 mm& 0.5 mm & $2.8$\\
Clouds vertical expansion vel. &3 mm/s& 0.3 mm/s & $<1$\\
Clouds transversal expansion vel.&15 mm/s& 2 mm/s & $4.2$\\
Launch direction change (Coriolis force) & - & $\sim$1 $\mu$rad & $3.6$\\
Magnetic fields & - & - & $<1$\\
Density homogeneity & $10^{-3}$ & $10^{-4}$ & $10$\\
Source mass geometry & - & 1 $\mu$m & $2.5$\\
\br
Total & & & $12$ \\
\br
\end{tabular}
\end{indented}
\end{table*}

Even if, compared to \cite{Rosi2014}, the selected uncertainties are equal or twice as large, the total systematic error remains below 10 ppm. As expected, the sensitivity with respect to the vertical coordinates is considerably suppressed. This can also be seen from the point of view of the mass positioning:
an error of $5$ mm induces a total shift of only 2 ppm. Therefore there is no need for micrometric translation stages and optical rulers, a valuable advantage when a heavy object must be translated over long distances ($\sim4$ m for Far configuration). Radially, the overall $G$ dependency is
reduced by a factor $\sim10$, much more than expected from the scaling factor of the apparatus.

\subsection{Other effects}
Approaching the 10 ppm accuracy requires further attention also to other parameters that did not constitute a limit in previous determinations. The most problematic, especially for a cold atomic fountain, is the Coriolis shift that can bias the interferometer reading if transversal velocity of the atomic samples and/or $k_{\mathrm{eff}}$ direction change when moving the source mass. Compensation schemes based on counteracting the
Earth's rotation rate by acting on the retroreflecting Raman mirror \cite{Hogan2007,Lan2012} must be optimized towards the percent accuracy. Moreover, the source mass movement system must be mechanically well isolated from the fountain holding structure in order to avoid correlations between source mass position and launch direction of the atomic sample. In this way such shift can be efficiently rejected and thus an improvement by a factor of 10 of the result reported in \cite{Rosi2014} appears feasible.

Magnetic fields must be kept under control along the interferometer region. An accurate design of the magnetic shields and the bias coils surrounding the interferometer tube is crucial to avoid external perturbations. However, it is worth remembering that modulating the signal by periodically switching between source mass configurations and reversing the $k_{\mathrm{eff}}$ vector \cite{Louchet-Chauvet2011} is a powerful strategy to further suppress these kind of shifts towards negligeble levels.

Another error source that is important to discuss arises from internal density inhomogeneities of the source mass. In order to make a rough evaluation, let us suppose to split our monolithic source mass into 100 thin disks (7.7 mm of thickness). Then we introduced a vertical square-wave density modulation, considering the central part of each disk (3.85 mm of thickness) less dense with respect the external ones. For a relative density variation of $(1.0\pm0.1)\times10^{-3}$ \cite{Lamporesi2007} and performing the usual Monte Carlo simulation we found a shift equal to ($-169\pm10$) ppm.

Finally, also the effect of variations in the source mass shape has been estimated. As anticipated in Section 4.1, a resolution of 1 $\mu$m on the source mass dimensions is required to ensure a negligeble shift of 2.5 ppm.

All the sources of errors discussed in this subsection have been added to table \ref{tab1}, yielding to a final projected uncertainty on $G$ of $12$ ppm.

\section{Conclusions and prospects}
In this paper a preliminary study of a determination of $G$ at $10$ ppm using a cold atom fountain is reported. With proper implementation of the method described in \cite{Roura2015}, systematic effects due to the cloud size, temperature and trajectories are suppressed.
In order to push the accuracy towards the $10^{-6}$ level, the implementation of ultra-cold atomic sources and large momentum transfer atom optics can be useful to enhance short term sensitivity and optimize the control over systematic shifts. If spurious magnetic fields represents the ultimate accuracy limit, using $^{88}$Sr instead of $^{87}$Rb could be a viable option \cite{Mazzoni2015}.
However, it is likely that main issues might come from the source mass itself. In this case shape characterization below $1$ $\mu$m seems unavoidable and inhomogeneities in the source mass material could be really hard to characterize.   

\ack
The author acknowledges L. Cacciapuoti and R. P. del Aguila for a critical reading of the manuscript.













%




%











%







\section*{References}

\bibliography{MAGIA2.bib}

\end{document}